\documentstyle[psfig,epsf]{mn}
\newcommand{\ltaraw}{$\; \buildrel < \over \sim \;$}
\newcommand{\lta}{\lower.5ex\hbox{\ltaraw}}
\newcommand{\gtaraw}{$\; \buildrel > \over \sim \;$}
\newcommand{\gta}{\lower.5ex\hbox{\gtaraw}}

\newcommand{\ffffff}[1]{\mbox{$#1$}}
\newcommand{\scnd}{\mbox{\ffffff{''}\hskip-0.3em.}}

\loadboldmathitalic 
\title [SCUBA observations of Hawaii 167]
{SCUBA observations of Hawaii 167} 
\author[G. F. Lewis \& S. C. Chapman]
{Geraint F. Lewis$^{1}$ \& S. C. Chapman$^{2}$\\
$^{1}$
Anglo-Australian Observatory, P.O. Box 296, Epping, NSW 1710, Australia:
Email \tt{gfl@aaoepp.aao.gov.au}\\
$^{2}$
Carnegie Observatories, 813 Santa Barbara Street, Pasadena, CA 91101, U.S.A.:
Email \tt{schapman@ociw.edu}
}
\date{\today}
\begin{document} 
\maketitle 
\begin{abstract}
We present  the first submillimetre  observations of the  z=2.36 broad
absorption  line  system  Hawaii~167.   Our observations  confirm  the
hypothesis that  Hawaii~167 contains a  massive quantity of  dust, the
optical  depth of  which is  sufficient to  completely  extinguish our
ultraviolet   view  of   a  central,   buried   quasar.   Hawaii~167's
submillimetre luminosity  and associated dust mass are  similar to the
ultraluminous class of infrared  galaxies, supporting the existence of
an evolutionary  link between these and the  active galaxy population.
Hawaii~167 appears  to be  a young quasar  which is emerging  from its
dusty cocoon.
\end{abstract}

\newcommand{\hawaii}{Hawaii~167}
\newcommand{\iras}{FSC~10214+4724}

\section{Introduction}\label{introduction}
\hawaii\ was identified  in the Hawaii K-band survey  (Songalia et al.
1994).   Displaying  a  rest-frame  ultraviolet spectrum  of  a  young
stellar population at z=2.34, \hawaii\ also possesses broad absorption
troughs  of both  high and  low ionized  species, consistent  with the
presence of bulk  outflows (Cowie et al.  1994).   While such features
are  characteristic  of  the  Broad  Absorption Line  (BAL)  class  of
quasars, the  broad emission lines indicative  of an AGN  core are not
seen  in the  ultraviolet spectrum.   Infrared  spectroscopy, however,
does reveal both broad ${\rm  H_\alpha}$ and ${\rm H_\beta}$ (Cowie et
al.  1994;  Egami et al.  1996),  with a redshift  of $z=2.36$.  These
observations  indicate  a substantial  Balmer-decrement  (Hall et  al.
1997), suggesting  that \hawaii\  is an example  of a  dust enshrouded
quasar; thought to represent an early evolutionary stage, such systems
would  appear as  fully fledged  quasars once  all the  dust  has been
removed.  During the transition from  one phase to another, as dust is
blown  from a central  obscuring torus,  the system  will appear  as a
BAL-type  quasar.  It  appears, therefore,  that \hawaii\  provides us
with a  view of  an embryonic quasar.   The limited data  available on
\hawaii, however,  means that an accurate determination  of dust mass,
total  luminosity and its  true evolutionary  status is  not currently
possible.

In  this  paper  we  present submillimetre  observations  of  \hawaii,
probing the emission  from a dusty component.  These  were obtained as
part  of a  survey of  the dust  properties of  BAL quasars  (Lewis \&
Chapman  2000  in  preparation).   In  Section~\ref{observations}  the
details     of    the     observations     are    presented,     while
Section~\ref{discussion} discusses  the dust content  and evolutionary
status of  \hawaii.  The  conclusions to this  study are  presented in
Section~\ref{conclusions}.

\section{Observations}\label{observations}
We  observed  \hawaii\ with  the  Submillimetre Common-User  Bolometer
Array (SCUBA) on the James Clerk Maxwell Telescope~\footnote{The James
Clerk Maxwell Telescope  is operated by The Joint  Astronomy Centre on
behalf of the  Particle Physics and Astronomy Research  Council of the
United Kingdom, the  Netherlands Organisation for Scientific Research,
and the  National Research Council  of Canada.} on Mauna  Kea, Hawaii.
We  used the  PHOTOMETRY three  bolometer chopping  mode  described in
Chapman et al.~(2000) and Scott et  al.~(2000) to keep the source in a
bolometer throughout  the observation.   This mode has  the additional
advantage of  allowing a check on  the apparent detection  of a source
over three independent bolometers.  The observations were taken in May
2000, using  the 450$\mu$m  and 850$\mu$m arrays  simultaneously.  The
alignment of the 450$\mu$m and 850$\mu$m arrays is not perfect, and we
did not  include the  450$\mu$m offbeams in  our final  flux estimate,
except to check  that the source had offbeam  flux consistent with the
detection in the primary bolometer.

For  our  double-difference  observations  there  are  instantaneously
$N\,{=}\,3$ beams, with the central beam having an efficiency of unity
and the two off beams having
\begin{equation}
\epsilon=-0.5\exp\left(-{d^2\over2\sigma_{\rm b}^2}\right),
\end{equation}
where  $d$ is the  angular distance  of the  off-beam centre  from the
source, and $\sigma_{\rm  b}$ is the Gaussian half-width  of the beam.
For   the  secondary   bolometer   the  beam   efficiency  is   simply
0.5. However,  distortion in  the field  results in  our  chosen third
bolometer being slightly offset from the source position, resulting in
a  beam  efficiency  of  0.44.   Our detection  level  increases  from
${\simeq}\,3.0\sigma$  to 3.7$\sigma$, after  folding in  the negative
flux density from the two offbeam pixels.

The effective integration time  on source was 1900\,ks.  The secondary
was chopped at  7.8125\,Hz with a chop throw at 53  arcsec to keep the
source on  bolometer at  all times.  Pointing  was checked  before and
after  the observation  on  blazars  and a  sky-dip  was performed  to
measure the atmospheric opacity directly. The rms pointing errors were
below  2 arcsec,  while the  average atmospheric  zenith  opacities at
450$\mu$m\ and  850$\mu$m\ were 1.5  and 0.22 respectively.   The data
were  reduced using the  Starlink package  SURF (Scuba  User Reduction
Facility, Jenness \& Lightfoot~1998) and our own reduction routines to
implement  the  three  bolometer  chopping mode.   Spikes  were  first
carefully  rejected   from  the  data,  followed   by  correction  for
atmospheric opacity  and sky subtraction  using the median of  all the
array pixels, except  for obviously bad pixels and  the source pixels.
The data  were then calibrated against standard  planetary and compact
\hbox{H\,{\sc   ii}~}  region  sources,   observed  during   the  same
night. The  resulting fluxes  were 6.04$\pm$1.65~mJy at  850$\mu$m and
66.0$\pm$20.7~mJy at 450$\mu$m.

\begin{figure*}
\centerline{ \psfig{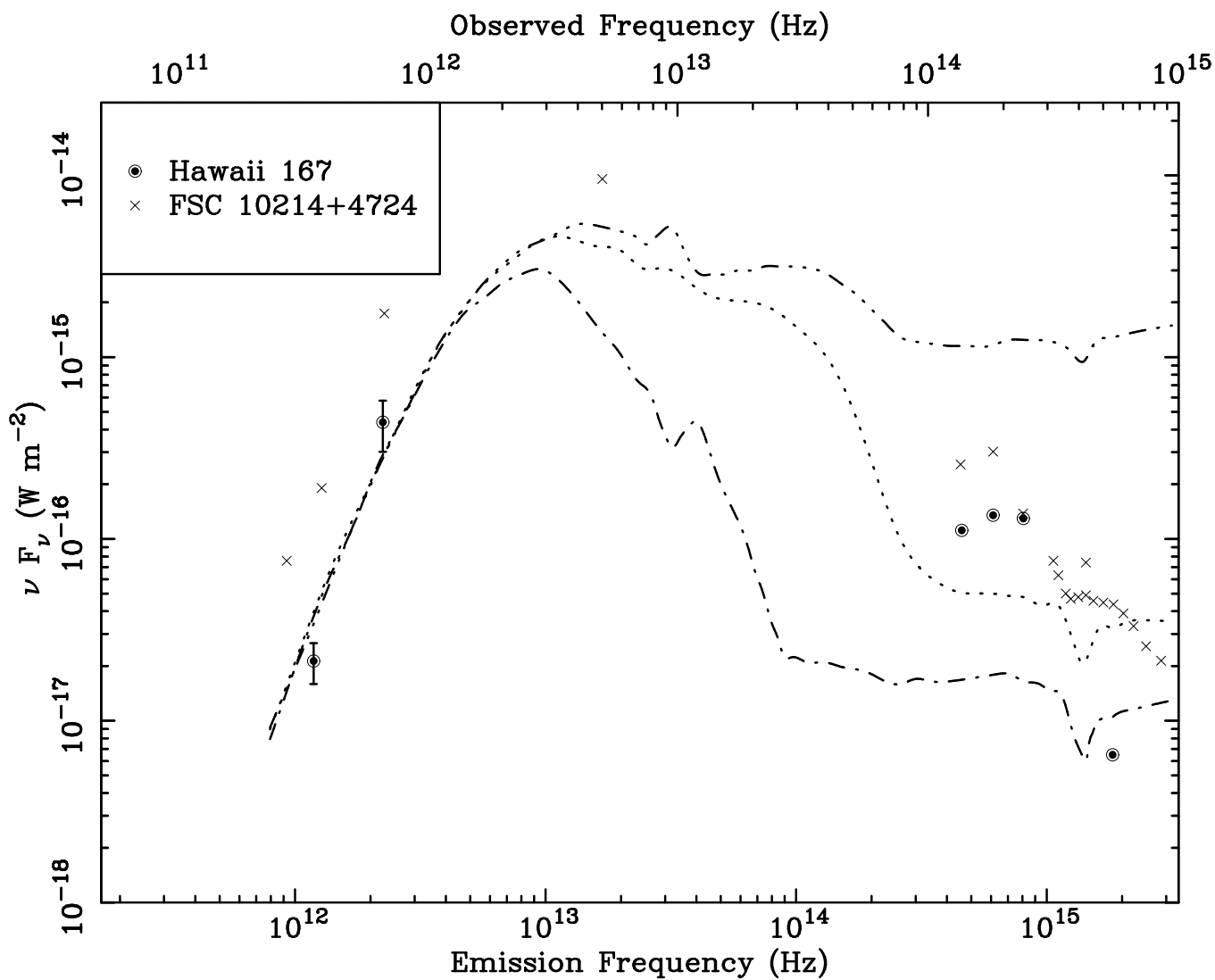} }
\caption{The spectral  energy distributions of  Hawaii~167 (z=2.36) as
compared  to  the  gravitationally  lensed, ultraluminous  galaxy  FSC
10214+4724   (z=2.28);  no   correction   for  gravitational   lensing
magnification has been  made to this data. The  data presented in this
paper are donated with error  bars.  The curves on each plot represent
the dust-enshrouded quasar models of  Granato et al.  (1996). From top
to  bottom,  these  represent  viewing  angles of  polar,  $45^o$  and
equatorial  respectively.   The   emission  frequency  corresponds  to
z=2.36.}
\label{SED}
\end{figure*}

\section{Discussion}\label{discussion}

\subsection{Luminosity and Dust Content}\label{dustcontent}
Figure~\ref{SED} presents  the extant  data of \hawaii,  including the
SCUBA data discussed in Section~\ref{observations}.  Also included for
comparison  is the spectral  energy distribution  (SED) of  the z=2.28
ultraluminous  infrared galaxy  \iras\ (Rowan-Robinson  et  al. 1993);
this system is also thought  to harbour a dust enshrouded quasar (e.g.
Barvainis  et  al.  1995).   Gravitational  lensing has  significantly
boosted the apparent  luminosity of \iras\ (e.g. by  30$\times$ in the
IR;   Broadhurst  \&   Lehar   1995),  although   no  correction   for
magnification has been made to its data points in the figure.

Emission from  a dusty component can  be modeled as a  greybody of the
form
\begin{equation}
{\rm
F_\nu \propto \frac{ \nu^3 }{exp^{\left(\frac{h\nu}{kT}\right)} - 1} 
\left[ 1 - exp^{ -\left(\frac{\nu}{\nu_o}\right)^\beta} \right]
}
\label{benford}
\end{equation}
where  ${\rm \nu_o}$  is the  frequency  at which  the source  becomes
optically thick  (Benford et al. 1999);  the shape of  the greybody is
only mildly sensitive  to the value of ${\rm \nu_o}$  and we adopt the
value of  ${\rm 2.4THz}$ (Hughes et  al. 1993).  The  \hawaii\ data is
consistent with a greybody temperature of ${\rm T=88\pm25K}$ and ${\rm
\beta=2.8\pm1.1} $; while these  values differ from `typical' greybody
fits  to  AGN, which  result  in values  of  ${\rm  T=50K}$ and  ${\rm
\beta=1.5}$ (Benford et al. 1999), the differences are not significant
given the errors.

Following the recipe of McMahon et al.  (1999), the 850$\mu$m flux was
use  to calculate  the  following physical  properties~\footnote{${\rm
(\Omega_o=1,    \Lambda_o=0,    H_o=50h_{50}\    km/s/Mpc}$    assumed
throughout${\rm   )}$};   a   far   infrared   luminosity   of   ${\rm
L_{FIR}\sim1.1\times10^{13}\  h_{50}^{-2} L_\odot}$ and  an associated
dust   mass  of   ${\rm  M_d=3.3\times10^{8}\   h_{50}^{-2}  M_\odot}$
(assuming  a dust  temperature of  50K).  Such  values  place \hawaii\
firmly  in the ultraluminous  class of  infrared galaxies  (Sanders \&
Mirabel 1996).  If  the observed infrared emission were  solely due to
stars,  it  would  correspond  to  a  star  formation  rate  of  ${\rm
\sim1000~M_\odot/yr}$ (this  is an upper limits  as \hawaii\ obviously
contains  a central  AGN source  which also  acts to  heat  the dust).
\hawaii\  displays  no evidence  of  gravitational lensing,  appearing
point-like at a PSF-scale $0\scnd5$ (Cowie et al. 1994), and hence its
inferred properties have  not been magnified and we  can conclude that
\hawaii\ is truly an ultraluminous system.
 
While we have  added to the SED of \hawaii, the  data are still sparse
and a detailed determination of the underlying physical properties and
geometry  is unwarranted.  Instead,  we compare  its SED  to published
models  of dust  enshrouded  quasars.  The  lines in  Figure~\ref{SED}
represent  such models for  ultraluminous infrared  galaxies (Granato,
Danese \& Franceseschini 1996).  Consisting of a optically thick torus
of dust,  extending over  several hundred parsecs  from the  AGN core,
each curve  represents a differing viewing angle;  the dot-dashed line
is  an equatorial  view  which maximally  extinguishes the  optical-UV
light from  the quasar,  which the treble-dot-dashed  line is  a polar
view  with an  unobstructed view  of the  AGN core.   The  dotted line
represents an intermediate case, with a viewing angle of $\sim45^o$. A
substantial  portion  of the  radiation  from  the  central quasar  is
reprocessed into submm/IR by dust in the torus. Such models reasonably
reproduce the SED characteristics of \iras\ (equatorial view), as well
as  other  ultraluminous  systems   such  as  the  cloverleaf  quasar,
H1413+117  (polar  view)  and IRAS~09104+4109  \&  IRAS~FSC~15307+3252
(intermediate  view).   In Figure~\ref{SED},  these  curves have  been
normalized to  the data presented in  this paper.  It  is important to
note  that Egami  et al.   (1996)  conclude that  the (rest-frame)  UV
emission in  \hawaii\ arises solely  in a stellar population,  with no
contribution  from the  AGN core  due to  complete obscuration  by the
circumnuclear dust. As all the models of Granato et al. (1996) predict
that {\it some} radiation from  the central regions must be visible in
the UV, none can accurately describe  the SED of \hawaii, but this may
be due to the parameter set employed in the modeling; the inclusion of
more dust, or changing the opening angle of the torus may bring better
agreement between the  models and the data.  It  is apparent, however,
that  combining the submillimetre  and infrared  data favours  a model
with an intermediate viewing angle  on to the obscuring torus.  Again,
scaling from the models, the  resulting dust mass in \hawaii\ is ${\rm
\sim  3\times 10^7\  M_\odot}$.  Without  more  data over  the SED  of
\hawaii,  this value  possesses  a significant  uncertainty, but  does
indicate that \hawaii\ harbours a vast quantity of dusty material.

\subsection{Broad Absorption Lines}\label{BAL}
The nature of the broad absorption  lines seen in the rest frame UV of
\hawaii\ presents  an interesting problem. In the  `standard model' of
BAL  quasars the  prominent absorption  lines  are the  result of  the
central continuum  emission being  observed through material  which is
ablated from  an obscuring torus by  the action of  the central quasar
(e.g. Barvainis et al.  1995). In  \hawaii\ the UV view of the central
quasar  is  completely  obscured  and  the  AGN  radiation  cannot  be
responsible for accelerating the  BAL material.  A potential solution,
however,  is that  the BAL  material is  driven from  the  outer dusty
regions by  hot young  stars. The requirement  of two sources  for the
driving  force of  the BAL  material  during different  stages of  the
systems evolution does,  however, seem a little contrived,  but we can
currently offer no solution to the problem.

\subsection{Evolutionary State and Further Study}\label{Evolution}
In  terms of  infrared  luminosity, and  hence  associated dust  mass,
\hawaii\  is similar  to other  ultraluminous  systems (Rowan-Robinson
2000), and  it is only  the detection of  broad emission lines  in the
infrared  that  directly reveals  the  presence  of  a quasar  at  its
core. Such  a picture is  consistent with their being  an evolutionary
link  between the  two populations  (e.g.  Sanders  \&  Mirabel 1996).
Here, an  initial burst of star  formation is triggered  by the merger
between  two  gas-rich systems.   The  merger  channels  gas into  the
central  regions  of  the  remnant,  forming  and  feeding  an  quasar
core. This AGN, however, is obscured by dust, the detritus of the star
burst; it is at this stage which we find \hawaii.  The dust is ablated
from  the torus  due to  radiation  from the  quasar core,  eventually
clearing and revealing an `normal' quasar.

Further clues to  the nature and geometry of  \hawaii\ will be gleaned
from polarization  studies.  As demonstrated with  \iras\ (Goodrich et
al. 1996),  such observations can  reveal the presence  broad emission
features, and hence an AGN core, the light of which has been scattered
from a exterior region (Barvainis  et al. 1995). The identification of
a scattered view of the central engine will imply that the AGN core is
not completely obscured.

\section{Conclusions}\label{conclusions}
We have  presented new submillimetre photometry of  the high redshift,
broad absorption line system \hawaii.  These observations confirm that
this  system consists of  a quasar  which is  enshrouded in  a massive
(${\rm \sim 10^7\rightarrow10^8\ M_\odot}$)  quantity of dust. With an
inferred       infrared      luminosity       of       ${\rm      \sim
10^{13}~h_{50}^{-2}~L_\odot}$,   \hawaii\   is   a   member   of   the
ultraluminous class of infrared  galaxies. The more extreme members of
this  family, namely  \iras\ and  H~1413+117,  have been  found to  be
gravitational  lenses. Hence,  their apparent  luminosities  have been
significantly  magnified.  Considering  this,  \hawaii\ represents  an
intrinsically more luminous source than these objects.

The  identification of  \hawaii\ as  an ultraluminous  infrared galaxy
provides more supporting evidence for there being an evolutionary link
between these  and the AGN family.  While presenting us  with the rare
view  of a  embryonic  quasar in  the  process of  shedding its  dusty
cocoon,  data  on  \hawaii\   are  currently  quite  sparse  and  more
observations are required before detailed modeling can be undertaken.

\section*{Acknowledgements}
We thank  the staff of  the JCMT for  their assistance with  the SCUBA
observations, and  the weather for  being so cooperative.   GFL thanks
the Australian Nuclear Science  \& Technology Organization (ANSTO) for
financial support.

\end{document}